\documentstyle[11pt]{article}

\begin{document}

\begin{center}
{\large {\bf Probability Amplitudes for Charge-Monopole Scattering}}
\end{center}

\begin{center}
{\large Saulo Carneiro}
\end{center}

\begin{center}
{\it Instituto de F\'{\i}sica, Universidade Federal da Bahia\\40210-340,
Salvador, BA, Brasil}
\end{center}

\begin{abstract}
In this letter we quantize a previously proposed non-local lagrangean for
the classical dual electrodynamics ({\it Phys.Lett.B 384(1996)197}), showing
how it can be used to construct probability amplitudes. Our results are
shown to agree with those obtained in the context of Schwinger and Zwanziger
formulations, but without necessity of introducing strings. \vspace{0.5cm}
\end{abstract}

Magnetic monopoles have played a remarkable role in particle
physics along the years, as elementary particles in abelian dual
electrodynamics$^{[1]}$ or topological solutions of non-abelian unified
theories$^{[2]}$, but a quantal approach to the interaction between charges and
poles has been always a challenging open problem, due to two different
difficulties$^{[3]}$: the non-perturbative character of the charge-pole
interaction and the absence of a complete lagrangean formulation, connected
to the impossibility of introducing regular $4$-potentials.

In a previous work$^{[4,5]}$, a covariant and gauge-invariant, manifestly
dual, non-local lagrangean formalism has been reported, leading to the
complete set of dual electromagnetic equations, without necessity of any
subsidiary condition or constraint on the particles motion. Now, dismissing
the problem of the non-perturbative value of the magnetic charge, we
quantize such an approach, constructing an invariant perturbative theory for
the charge-monopole interaction.

The referred non-local lagrangean, obeying a saddle-point action principle,
has the interaction sector

\begin{equation}  \label{1}
{\cal L}_{int} = - j_{\mu} {\cal A}^{\mu} + g_{\mu} \tilde{{\cal A}}^{\mu}
\end{equation}

\noindent where $j^{\mu}$ and $g^{\mu}$ are, respectively, the electric and
magnetic $4$-currents, and the non-local potentials ${\cal A}^{\mu}$ and $%
\tilde{{\cal A}}^{\mu}$ are defined by

\begin{equation}  \label{2}
{\cal A}^{\mu} = A^{\mu} + \frac{1}{2} \epsilon^{\mu \gamma \alpha \beta}
\int_P^x \partial_{\alpha} \tilde{A}_{\beta} \; d\xi_{\gamma}
\end{equation}

\begin{equation}  \label{3}
\tilde{{\cal A}}^{\mu} = \tilde{A}^{\mu} - \frac{1}{2} \epsilon^{\mu \gamma
\alpha \beta} \int_{\tilde{P}}^x \partial_{\alpha} A_{\beta} \; d\xi_{\gamma}
\end{equation}

Here, $A^{\mu}$ and $\tilde{A}^{\mu}$ are the local potentials of Cabibbo
and Ferrari$^{[6]}$, defined, in terms of the field strength, by

\begin{equation}  \label{4}
F^{\mu \nu} = \partial^{\mu} A^{\nu} - \partial^{\nu} A^{\mu} -
\epsilon^{\mu \nu \alpha \beta} \; \partial_{\alpha} \tilde{A}_{\beta}
\end{equation}

\noindent and obeying, in Lorenz's gauge, the wave equations$^{[7]}$

\begin{equation}  \label{4'}
\partial^{\nu} \partial_{\nu} A^{\mu} = -j^{\mu}
\end{equation}

\begin{equation}  \label{4''}
\partial^{\nu}\partial_{\nu} \tilde{A}^{\mu} = -g^{\mu}
\end{equation}

The interaction operator corresponding to $(\ref{1})$ is given by$^{[4]}$

\begin{equation}  \label{5}
V = - \frac{\partial S^e_{int}}{\partial t} + \frac{\partial S^g_{int}}{%
\partial t} = \int d^3x \; (j_{\mu} {\cal A}^{\mu} + g_{\mu} \tilde{{\cal A}}%
^{\mu})
\end{equation}

\noindent where $S^{e}_{int}$ and $S^{g}_{int}$ stand for the charge and monopole
interaction actions, respectively. So we have, for the scattering matrix,

\begin{equation}  \label{6}
S = T exp \left\{ -i \int d^4x \; (j_{\mu} {\cal A}^{\mu} + g_{\mu} \tilde{%
{\cal A}}^{\mu}) \right\}
\end{equation}

Expanding it in powers of the electric and magnetic charges, $e$ and $g$, the
first non-diagonal contribution is the second order one

\begin{eqnarray}  \label{7}
S^{(2)} &=& - \frac{1}{2} \int \int d^4x \; d^4x^{\prime}\; \{ T[j^{\mu}(x)
j^{\nu}(x^{\prime})] T[{\cal A}_{\mu}(x) {\cal A}_{\nu}(x^{\prime})] + \\
&+& T[g^{\mu}(x) g^{\nu}(x^{\prime})] T[\tilde{{\cal A}}_{\mu}(x) \tilde{%
{\cal A}}_{\nu}(x^{\prime})] + 2 T[j^{\mu}(x) g^{\nu}(x^{\prime})] T[{\cal A}%
_{\mu}(x) \tilde{{\cal A}}_{\nu}(x^{\prime})]\}  \nonumber
\end{eqnarray}

The first and second terms correspond, respectively, to charge-charge and
pole-pole scatterings. In the first case, we can use gauge invariance to put 
$\tilde{A}^{\mu} = 0^{[9]}$ and to introduce the photon propagation function

\begin{equation}  \label{8}
D_{\mu \nu}(x-x^{\prime}) = i <0|T{\cal A}_{\mu}(x) {\cal A}%
_{\nu}(x^{\prime})|0> = i <0|TA_{\mu}(x) A_{\nu}(x^{\prime})|0>
\end{equation}

\noindent and, in the second case, putting $A^{\mu} = 0$ we can use the
propagation function

\begin{eqnarray}  \label{9}
\tilde{D}_{\mu \nu}(x-x^{\prime}) = i <0|T\tilde{{\cal {A}}}_{\mu}(x) \tilde{%
{\cal {A}}}_{\nu}(x^{\prime})|0> = i <0|T\tilde{A}_{\mu}(x) \tilde{A}%
_{\nu}(x^{\prime})|0>
\end{eqnarray}

\noindent leading to a pole-pole interaction completely analogous to the
charge-charge one.

The last term in $(\ref{7})$ corresponds to the charge-monopole scattering
and suggests the introduction of the mixed propagation function

\begin{equation}  \label{10}
C_{\mu \nu}(x-x^{\prime}) = 2 i <0|T{\cal A}_{\mu}(x) \tilde{{\cal A}}%
_{\nu}(x^{\prime})|0>
\end{equation}

\noindent This allows us to write the scattering amplitude in the form

\begin{equation}  \label{11}
M = eg (\bar{u}_g \gamma^{\mu} u_g) C_{\mu \nu} (\bar{u}_e \gamma^{\nu} u_e)
\end{equation}

\noindent where $u_g$ and $u_e$ are, respectively, the pole and charge
amplitudes\footnote{%
To fix ideas, we are considering charge and pole as $1/2$-spin particles.}.

It is easy to see that $<0|TA_{\mu}(x) \tilde{A}_{\nu}(x^{\prime})|0> = 0$,
due to the fact that $A^{\mu}$ and $\tilde{A}^{\mu}$ describe photons with
opposite parities\footnote{%
Actually, this reasoning is not necessary: by fixing the gauges $A^{\mu}=0$
or $\tilde{A}^{\mu}=0$, this expectation value vanishes trivially.}. Thus,
using $(\ref{2})$ and $(\ref{3})$ we obtain, from $(\ref{10})$,

\begin{eqnarray}  \label{12}
C_{\mu \nu}(x-x^{\prime}) &=& i \epsilon_{\mu \gamma \alpha \beta} <0|T%
\tilde{A}_{\nu}(x^{\prime})\int_P^x \partial^{\alpha} \tilde{A}^{\beta} \;
d\xi^{\gamma}|0> - \\
& & - i \epsilon_{\nu \gamma \alpha \beta} <0|TA_{\mu}(x)\int_{\tilde{P}%
}^{x^{\prime}} \partial^{\alpha} A^{\beta} \; d\xi^{\gamma}|0>  \nonumber
\end{eqnarray}

Remembering that, in the classical limit, $P$ ($\tilde{P}$) coincides with
the charge (pole) world-line between $\xi = - \infty$ and $\xi = x$ ($%
x^{\prime}$), it is a straightforward calculation to verify that, in $(\ref
{12})$, the chronological ordering operator $T$ commutes with the integral
and derivative operators. Then, we have

\begin{eqnarray}  \label{13}
C_{\mu \nu}(x-x^{\prime}) &=& i \epsilon_{\mu \gamma \alpha \beta} \int_P^x
d\xi^{\gamma} \; \partial^{\alpha}_{\xi} <0|T\tilde{A}_{\nu}(x^{\prime}) 
\tilde{A}^{\beta}(\xi)|0> - \\
& & - i \epsilon_{\nu \gamma \alpha \beta} \int_{\tilde{P}}^{x^{\prime}}
d\xi^{\gamma} \; \partial^{\alpha}_{\xi} <0|TA_{\mu}(x) A^{\beta}(\xi)|0> 
\nonumber \\
&=& \epsilon_{\mu \gamma \alpha \beta} \int_P^x d\xi^{\gamma} \;
\partial^{\alpha}_{\xi} \tilde{D}^{\beta}_{\nu}(x^{\prime}-\xi) -
\epsilon_{\nu \gamma \alpha \beta} \int_{\tilde{P}}^{x^{\prime}}
d\xi^{\gamma} \; \partial^{\alpha}_{\xi} D^{\beta}_{\mu}(x-\xi)  \nonumber \\
&=& \epsilon_{\mu \gamma \alpha \beta} \int_P^{x-x^{\prime}} d\xi^{\gamma}
\; \partial^{\alpha}_{\xi} \tilde{D}^{\beta}_{\nu}(\xi) - \epsilon_{\nu
\gamma \alpha \beta} \int_{\tilde{P}}^{x^{\prime}-x} d\xi^{\gamma} \;
\partial^{\alpha}_{\xi} D^{\beta}_{\mu}(\xi)  \nonumber
\end{eqnarray}

Thus, we can write the mixed propagator in terms of the propagators $(\ref{8}%
)$ and $(\ref{9})$, as

\begin{eqnarray}  \label{14}
C_{\mu \nu}(x) = \epsilon_{\mu \gamma \alpha \beta} \int_P^{x} d\xi^{\gamma}
\; \partial^{\alpha}_{\xi} \tilde{D}^{\beta}_{\nu}(\xi) - \epsilon_{\nu
\gamma \alpha \beta} \int_{\tilde{P}}^{-x} d\xi^{\gamma} \;
\partial^{\alpha}_{\xi} D^{\beta}_{\mu}(\xi)
\end{eqnarray}

Once more, we can use the gauges $A_{\mu}=0$ or $\tilde{A}_{\mu}=0$ in order
to obtain, respectively,

\begin{eqnarray}  \label{14'}
C_{\mu \nu}(x) = \epsilon_{\mu \gamma \alpha \beta} \int_P^{x} d\xi^{\gamma}
\; \partial^{\alpha}_{\xi} \tilde{D}^{\beta}_{\nu}(\xi)
\end{eqnarray}

\noindent and

\begin{eqnarray}  \label{14''}
C_{\mu \nu}(x) = - \epsilon_{\nu \gamma \alpha \beta} \int_{\tilde{P}}^{-x}
d\xi^{\gamma} \; \partial^{\alpha}_{\xi} D^{\beta}_{\mu}(\xi)
\end{eqnarray}

We see that the mixed propagator (and then the amplitude $(\ref{11})$) is a
non-local quantity, depending on the integration paths $P$ and $\tilde{P}$.
Nevertheless, like in the classical case, this non-locality will proven to be non-observable when we calculate observable quantities like $|M|^2$ and the
scattering cross section.

Indeed, these observables can be obtained if we calculate, from (\ref{14'}),
the local quantity\footnote{%
See footnote $2$ in $[4]$.}

\begin{equation}  \label{15}
\partial_{\lambda} C_{\mu \nu}(x) = \epsilon_{\mu\lambda \alpha\beta}
\partial^{\alpha} \tilde{D}^{\beta}_{\nu}(x)
\end{equation}

Then we see that the mixed propagator obeys the equation

\begin{equation}
\partial^{\lambda}\partial_{\lambda}C_{\mu\nu} = \epsilon_{\mu\nu\lambda
\alpha}[\partial^{\lambda},\partial^{\alpha}]D_F(x)
\end{equation}

\noindent where we have used $\tilde{D}^{\beta\nu}(x)=g^{\beta\nu}D_F(x)$ 
\footnote{
It is important to remark that the commutator $[\partial^{\lambda},
\partial^{\alpha}]D_F(x)$ does not vanish in the whole space-time, due to the
discontinuity of the Feynman propagator $D_F(x)$ at $t=0$.}.

In the momentum representation, (\ref{15}) has the form

\begin{equation}  \label{16}
k_{\lambda} C_{\mu \nu}(k) = \epsilon_{\mu \lambda \alpha \beta} k^{\alpha} 
\tilde{D}^{\beta}_{\nu}(k)
\end{equation}

Using for the photon propagator

\begin{equation}  \label{17}
\tilde{D}^{\mu \nu}(k) = \frac{4\pi}{k^2} g^{\mu \nu}
\end{equation}

\noindent we have

\begin{eqnarray}  \label{18}
k_{\lambda} C_{\mu \nu}(k) = \frac{4\pi}{k^2} k^{\alpha} \epsilon_{\mu \nu
\lambda \alpha}
\end{eqnarray}

It is easy to show that using (\ref{14''}) leads to the same result, which means that the exchange of the two kinds of photons present in the theory (the
fields $A^{\mu}$ and $\tilde{A}^{\mu}$) gives identical contributions to
observable quantities, the two kinds of photons being indistinguishable from
the observational point of view. This is related to the fact that, in the
classical version of the theory, introducing the additional $4$-potential $%
\tilde{A}^{\mu}$ does not change the number of independent physical degrees
of freedom, due to the presence of extra gauge invariance$^{[9]}$.

>From $(\ref{18})$ and $(\ref{11})$, we can derive

\begin{eqnarray}  \label{19}
k_{\lambda} M = eg (\bar{u}_g \gamma^{\mu} u_g) k_{\lambda} C_{\mu \nu} (%
\bar{u}_e \gamma^{\nu} u_e) = \frac{4\pi eg}{k^2} k^{\alpha} \epsilon_{\mu
\nu \lambda \alpha} (\bar{u}_g \gamma^{\mu} u_g) (\bar{u}_e \gamma^{\nu} u_e)
\end{eqnarray}

Therefore,

\begin{equation}  \label{20}
|k_{\lambda} M|^2 = k^2 |M|^2 = \frac{32e^2g^2\pi^2}{k^4} U_{\mu \nu}
U^{\dag}_{\alpha \beta} k_{\sigma} k_{\delta} \epsilon^{\mu \nu \lambda
\sigma} \epsilon_{\lambda}^{\; \; \alpha \beta \delta}
\end{equation}

\noindent or

\begin{equation}  \label{21}
|M|^2 = \frac{32e^2g^2\pi^2}{k^6} U_{\mu \nu} U^{\dag}_{\alpha \beta}
k_{\sigma} k_{\delta} \epsilon^{\mu \nu \lambda \sigma}
\epsilon_{\lambda}^{\; \; \alpha \beta \delta}
\end{equation}

\noindent where we have introduced

\begin{equation}  \label{22}
U^{\alpha \beta} \equiv (\bar{u}_g \gamma^{\alpha} u_g) (\bar{u}_e
\gamma^{\beta} u_e)
\end{equation}

>From $(\ref{21})$, we see that only the antisymmetric part of $U_{\mu \nu }$
and $U_{\alpha \beta }^{\dag }$ will contribute to $|M|^2$. A direct
calculation leads us to

\begin{equation}  \label{23}
\frac{1}{64\pi^2} |M|^2 = \frac{e^2g^2}{k^4} U^{[\mu \nu]}
U^{\dag}_{[\mu\nu]}
\end{equation}

\noindent where $U^{[\mu \nu ]}$ indicates the antisymmetric part of $U^{\mu
\nu }$, and where we have used the transversality conditions $k^\mu j_\mu
=k^\mu g_\mu =0$, expressing the conservation of the electric and magnetic
currents in the momentum representation.

At this point it is interesting to compare our results with those obtained
by Dirac-type string formulations. On the basis of Schwinger's formalism$%
^{[10]}$, Rabl$^{[11]}$ derived the gauge-dependent propagator

\begin{equation}  \label{Rabl1}
C^{\prime}_{\mu\nu}(k) = \frac{4\pi}{k^2} \frac{\epsilon_{\mu\nu\lambda%
\alpha}n^{\lambda}k^{\alpha}}{(n \cdot k)}
\end{equation}

\noindent where $n^{\lambda}$ is a fixed but arbitrary $4$-vector pointing
in the string direction. This propagator was also obtained by Zwanziger$%
^{[12]}$ in the context of another string-based formulation\footnote{%
For a good review on the Schwinger and Zwanziger approaches, see $[3]$.}.

Equation (\ref{Rabl1}) can be rewritten as

\begin{equation}  \label{Rabl2}
n^{\lambda} (k_{\lambda}C^{\prime}_{\mu\nu}) = n^{\lambda} \left(\frac{4\pi}{%
k^2} k^{\alpha} \epsilon_{\mu\nu\lambda\alpha}\right)
\end{equation}

\noindent And, as $n^{\lambda}$ is an arbitrary $4$-vector, we have

\begin{equation}  \label{Rabl3}
k_{\lambda}C^{\prime}_{\mu\nu} = \frac{4\pi}{k^2} k^{\alpha}
\epsilon_{\mu\nu\lambda\alpha}
\end{equation}

\noindent which must be compared to (\ref{18}).

On the other hand, multiplying (\ref{18}) by an arbitrary $4$-vector $%
n^{\lambda}$, we can put it in the form

\begin{equation}  \label{Rabl4}
C_{\mu\nu}(k) = \frac{4\pi}{k^2} \frac{\epsilon_{\mu\nu\lambda\alpha}
n^{\lambda}k^{\alpha}}{(n \cdot k)}
\end{equation}

>From (\ref{Rabl1}) and (\ref{Rabl4}) we see that $C_{\mu\nu} =
C^{\prime}_{\mu\nu}$, that is, our mixed propagator is equal to the
propagator derived in the Schwinger and Zwanziger formulations.

While equation (\ref{14}) exhibits the non-local character of the mixed
propagator, equation (\ref{Rabl4}) (which depends on the arbitrary $4$%
-vector $n^\lambda $) shows its gauge-dependence$^{[3]}$. Actually, the
relation between non-locality and gauge-dependence is already present in the
classical version of the theory: a change of the paths of integration $P$
and $\tilde P$ leads only to a gauge transformation of the non-local
potentials (\ref{2}) and (\ref{3})$^{[13]}$. By the way, let us emphasize
that it is this property that guarantees the full covariance of the
formalism and the strict locality of observables and equations of motion.

Let us calculate the classical limit of the probability (\ref{23}),
obtaining the differential cross section for the elastic, non-relativistic,
scattering of an electron by a massive monopole at rest. In the classical
limit, the electric and magnetic $4$-currents are given by

\begin{equation}
j^{\mu}=e\bar{u}_e\gamma^{\mu}u_e=e\gamma_e(1,\vec{v}_e)
\end{equation}

\begin{equation}
g^{\mu}=g\bar{u}_g\gamma^{\mu}u_g=g\gamma_g(1,\vec{v}_g)
\end{equation}

\noindent where $\gamma_{e,g}=(1-v^2_{e,g})^{-1/2}$ stand for the Lorentz
factors for the charge and pole, respectively.

By using these currents in (\ref{23}) we obtain, in the monopole rest frame,

\begin{equation}
\frac{1}{16\pi^2} |M|^2 = \frac{e^2g^2v_e^2\gamma_e^2}{k^4}
\end{equation}

\noindent which leads, in the non-relativistic limit, for a small angle of
scattering, to the differential cross section

\begin{equation}  \label{goldhaber}
d\sigma=\frac{e^2g^2v_e^2}{|\vec{p}|^4\theta^4}d\Omega
\end{equation}

\noindent where $\vec{p}$ is the momentum of, say, the incident electron.

This result was originally obtained by Goldhaber in the context of a
non-relativistic approach$^{[11,14]}$. It can be interpreted as the cross
section for the Coulomb scattering of a charge $e$ by a charge $gv_e$, in
accordance with the fact that, in the classical theory, a static charge-pole
pair does not interact. By the way, let us note that the dependence on the
relative velocity explains the obtainment of (\ref{goldhaber}) in the
context of a perturbative expansion, despite magnetic charge being
non-perturbative: in this non-relativistic limit, the effective charge $%
gv_e<<1$.

Finally, let us briefly comment the question of {\it dyons}. The formalism
proposed in $[4]$ is invariant only under the discrete dual transformation
corresponding to a dual angle of $\pi /2$, i.e., the transformation that
interchanges the electric and magnetic charges; it is not invariant under a
general dual transformation with arbitrary dual angle, what means in
particular that our lagrangean is not appropriate to describe elementary
dyons. This is intimately connected to the saddle-point character of the
action on which the formalism is based: an elementary particle cannot
simultaneously minimize (as an electric charge) and maximize (as a
monopole) the action. Of course nothing forbids one to describe dyons as
composite systems but, if we want to describe them as elementary ones, some
generalization of the theory is needed.

\section*{Acknowledgments}

I would like to thank Drs. E. Ferrari, J. Frenkel, J.A. Helayel-Neto, E.
Recami and F. Rohrlich for useful discussions and suggestions. My thanks
also to Dr. C.O. Escobar for the hospitality in {\it Universidade de S\~ao
Paulo} during the elaboration of this paper, and to Prof. N. Andion for the
reading of the manuscript.

\begin{eqnarray}
h_{\mu}(x) = \frac{1}{2} \epsilon_{\mu \gamma \alpha \beta} \int d^4y \;
g^{\gamma}(y) \; \left[\int_{\tilde{P}}^y \partial^{\alpha}_x
\delta^4(x-\xi) d\xi^{\beta}\right]  \nonumber
\end{eqnarray}

Nevertheless, it is possible to verify that such extra terms vanish identically. Indeed, integrating by parts, we can rewrite the above quantity as%
\footnote{%
See footnote $2$ in $[4]$.}

\begin{eqnarray}
h_{\mu}(x) &=& \frac{1}{2} \epsilon_{\mu\gamma\alpha\beta} \int
dS^{\beta\lambda\sigma} \; \left[\int_{\tilde{P}_S}^y \partial^{\alpha}_x
\delta^4(x-\xi) d\xi_{\lambda}\right] \; \left[\int_{\tilde{P}_S}^y
g^{\gamma}(\chi) d\chi_{\sigma}\right]  \nonumber \\
& & - \frac{1}{2} \epsilon_{\mu\gamma\alpha\beta} \int d^4y \;
\left[\partial^{\alpha}_x \delta^4(x-y)\right] \; \left[\int_{\tilde{P}}^y
g^{\gamma}(\xi) d\xi^{\beta}\right]  \nonumber
\end{eqnarray}

\noindent being $\tilde{P}_S$ on the hypersurface at infinity. Clearly, the
hypersurface term vanishes for any finite $x$ and so it does not contribute
to the action variation. Using $\int f(x) \delta^{\prime}(x) dx = - \int
\delta(x) f^{\prime}(x) dx$, the remaining term can be written as

\begin{eqnarray}
h_{\mu}(x) &=& - \frac{1}{2} \epsilon_{\mu\gamma\alpha\beta} \int d^4y \;
\delta^4(x-y) \; \left[\partial^{\alpha}_y \int_{\tilde{P}}^y
g^{\gamma}(\xi) \; d\xi^{\beta}\right]  \nonumber \\
&=& - \frac{1}{2} \epsilon_{\mu\gamma\alpha\beta} \; \delta^{\alpha\beta}
\int d^4y \; g^{\gamma}(y) \; \delta^4(x-y)  \nonumber \\
&=& - \frac{1}{2} \epsilon_{\mu\gamma\alpha\beta} \; \delta^{\alpha\beta} \;
g^{\gamma}(x) \equiv 0  \nonumber
\end{eqnarray}

\noindent due to the antisymmetry of $\epsilon _{\mu \gamma \alpha \beta }$.

I would like to thank Dr. N. Berkovits for interesting discussions about
this point.

\bibitem{8} N Berkovits, Phys.Lett.B 395(1997)28.

\bibitem{9} The field strength (\ref{4}) is invariant under the generalized gauge
transformations

\begin{eqnarray}
A_{\mu} \rightarrow A_{\mu} + A^{\prime}_{\mu}  \nonumber
\end{eqnarray}
\begin{eqnarray}
\tilde{A}_{\mu} \rightarrow \tilde{A}_{\mu} + \tilde{A}^{\prime}_{\mu} 
\nonumber
\end{eqnarray}

\noindent provided $A^{\prime}_{\mu}$ and $\tilde{A}^{\prime}_{\mu}$ satisfy
the zero field condition

\begin{eqnarray}
\partial^{\mu} A^{\prime \nu} - \partial^{\nu} A^{\prime \mu} -
\epsilon^{\mu \nu \alpha \beta} \; \partial_{\alpha} \tilde{A}
^{\prime}_{\beta} = 0  \nonumber
\end{eqnarray}

These gauge transformations generalize the usual ones

\begin{eqnarray}
A_{\mu} \rightarrow A_{\mu} + \partial_{\mu}\Lambda  \nonumber
\end{eqnarray}
\begin{eqnarray}
\tilde{A}_{\mu} \rightarrow \tilde{A}_{\mu} + \partial_{\mu}\Gamma  \nonumber
\end{eqnarray}

\noindent for which the zero field condition is identically satisfied. As
discussed in references $[3]$ and $[6]$, due to this extended gauge
invariance, the introduction of the new potential $\tilde{A}_{\mu}$ does not
increase the number of independent physical degrees of freedom of the
electromagnetic field.

\bibitem{10} J Schwinger, Phys.Rev. 144(1966)1087.

\bibitem{11} A Rabl, Phys.Rev. 179(1969)1363.

\bibitem{12} D Zwanziger, Phys.Rev.D 3(1971)880.

\bibitem{13} In fact, let us consider a magnetic monopole at rest in the origin of a
coordinate system (the electric charge distribution can be arbitrary). The
field equation (\ref{4''}) will have the Coulomb solution

\begin{eqnarray}
\vec{\tilde{A}}=0  \nonumber
\end{eqnarray}
\begin{eqnarray}
\tilde{A}^0=\frac{g}{4 \pi r}  \nonumber
\end{eqnarray}

Substituting this solution into (\ref{2}) we obtain

\begin{eqnarray}
{\cal A}^0 = A^0  \nonumber
\end{eqnarray}
\begin{eqnarray}  \label{vp}
\vec{{\cal A}}(\vec{r}) = \vec{A}(\vec{r}) - \frac{g}{8\pi}\int_P^{\vec{r}}%
\vec{\nabla}\left(\frac{1}{r^{\prime}}\right)\times d\vec{r}^{\prime} 
\nonumber
\end{eqnarray}

If we take a different integration path $P^{\prime}$, the new vector
potential $\vec{{\cal A}}^{\prime}$ will differ from $\vec{{\cal A}}$ by

\begin{eqnarray}
\vec{{\cal A}}^{\prime}(\vec{r}) - \vec{{\cal A}}(\vec{r}) = \frac{g}{8\pi}%
\oint_{P-P^{\prime}}\vec{\nabla}\left(\frac{1}{r^{\prime}}\right)\times d%
\vec{r}^{\prime}= \frac{g}{8\pi}\vec{\nabla}\Omega_0(\vec{r})  \nonumber
\end{eqnarray}

\noindent where $\Omega_0(\vec{r})$ is the solid angle formed by the contour 
$P-P^{\prime}$ in the origin. So, the two considered non-local potentials
differ by the gradient of a scalar function. Using now the covariance and
linearity of (\ref{2}), (\ref{4'}) and (\ref{4''}) we see that the $4$%
-potentials ${\cal A}^{\prime}_{\mu}$ and ${\cal A}_{\mu}$ differ by a gauge
transformation in any frame of reference and whatever the distribution of
charges and poles, as we wished to prove. This line of reasoning can also be
applied to the dual non-local potential $\tilde{{\cal A}}_{\mu}$.

\bibitem{14} AS Goldhaber, Phys.Rev. 140(1965)B1407.

\end{document}